%% file: lc95.tex
\documentstyle[sprocl,axodraw,epsfig]{article}
\def\d{\,{\rm d}}
\def\bcn{\begin{center}}
\def\ecn{\end{center}}

\newcommand{\lk}{leptoquark}
\newcommand{\bea}{\begin{eqnarray}}
\newcommand{\eea}{\end{eqnarray}}
\newcommand{\beq}{\begin{equation}}
\newcommand{\eeq}{\end{equation}}

\newcommand{\lc}{linear collider}
\newcommand{\sm}{standard model}

\newcommand{\xs}{cross section}

\newcommand{\EM}{electromagnetic}

\newcommand{\pe}{\mbox{$e^+e^-$}}

\newcommand{\ep}{\mbox{$e^-\gamma$}}

\def\lr3{$SU(3)_L\otimes SU(3)_R$}

\def\z0{$Z^0$}
\def\Z0{$Z^0$}

\def\gsim{\buildrel{\lower.7ex\hbox{$>$}}\over{\lower.7ex\hbox{$\sim$}}}
\def\lsim{\buildrel{\lower.7ex\hbox{$<$}}\over{\lower.7ex\hbox{$\sim$}}}
\begin{document}

\title{\boldmath 
LEPTOQUARKS IN \ep\ COLLISIONS
\unboldmath}

\author{FRANK CUYPERS}

\address{{\tt cuypers@pss058.psi.ch}\\
	Max-Planck-Institut f\"ur Physik,
	F\"ohringer Ring 6,
	D-80805 M\"unchen,
	Germany}


\maketitle\abstracts{
We discuss the \lk\ discovery potential 
of \ep\ scattering
and show how polarization and angular distributions
can be used to differentiate the different types of \lk s.
}

\noindent
The best present bounds on leptoquarks
still originate from low energy experiments~\cite{sacha},
and the constraints on flavour diagonal couplings are weak, 
at best.
Some improvement is expected from HERA~\cite{brw},
but the real breakthrough should be obtained
in high energy experiments of the next generation~\cite{br}.
A particularly promising option
is a \lc\ operated in the $e^-\gamma$ mode~\cite{eg},
where leptoquarks can be produced singly
and very well studied.

Since \lk s appear in a wealth of extensions of the \sm,
ranging from grand unified to composite models,
it is important to perform an analysis 
which is as model-independent as possible.
A general classification of these states,
which respects $SU(3)_c \otimes SU(2)_L \otimes U(1)_Y$ invariance,
was performed in Ref.~\cite{brw}.
It assumes that,
by definition,
they couple to leptons and quarks,
and must, therefore, be either singlets, doublets or triplets
of the weak gauge group $SU(2)_L$.
Also, for simplicity,
only lowest dimension operators are involved,
hence their couplings to leptons and quarks
do not involve derivatives.
Finally,
in order not to induce rapid proton decay 
or other nuisances,
they must conserve lepton $(L)$ and baryon $(B)$ number separately.
The most general effective lagrangian
which respects these conditions
reads~\cite{brw}
\begin{eqnarray}
\label{lag}
L
& = & 
\bigl( 
  h_{2L} \bar{u}_R \ell_L
+ h_{2R} \bar{q}_L i \sigma_2 e_R
\bigr) R_2
+ \tilde{h}_{2L} \bar{d}_R \ell_L \tilde{R}_2
+ h_{3L} \bar{q}_L \mbox{\boldmath$\sigma$} \gamma^\mu \ell_L \mbox{\boldmath$U$}_{3 \mu}
\\
& + &
\bigl( 
  h_{1L} \bar{q}_L \gamma^\mu \ell_L
+ h_{1R} \bar{d}_R \gamma^\mu e_R 
\bigr) U_{1 \mu}
+ \tilde{h}_{1R} \bar{u}_R \gamma^\mu e_R \tilde{U}_{1 \mu}
\nonumber\\
& + & 
\bigl( 
  g_{1L} \bar{q}^c_L i\sigma_2 \ell_L
+ g_{1R} \bar{u}^c_R e_R
\bigr) S_1
+ \tilde{g}_{1R} \bar{d}^c_R e_R \tilde{S}_1
+ g_{3L} \bar{q}^c_L i\sigma_2\mbox{\boldmath$\sigma$} \ell_L \mbox{\boldmath$S$}_3
\nonumber\\
& + & 
\bigl( 
  g_{2L} \bar{d}^c_R \gamma^\mu \ell_L
+ g_{2R} \bar{q}^c_L \gamma^\mu e_R
\bigr) V_{2 \mu}
+ \tilde{g}_{2L} \bar{u}^c_R \gamma^\mu \ell_L \tilde{V}_{2 \mu}
+ \mbox{ h.c.} \nonumber~,
\end{eqnarray}
where the \boldmath$\sigma$'s\unboldmath\ are Pauli matrices,
while $q_L$ and $\ell_L$ are the $SU(2)_L$ quark and lepton doublets
and $u_R$, $d_R$, $\ell_R$ are the corresponding singlets.
The subscripts of the \lk s
indicates the size of the $SU(2)_L$ representation
to which they belong.
The $R$- and $S$-type \lk s are spacetime scalars,
whereas the $U$ and $V$ are vectors.
All family and colour indices are implicit.

Since all these \lk s carry an electric charge,
they must also couple to the photon.
These interactions are described by the kinetic lagrangians 
for scalar and and vector bosons
\begin{eqnarray}
{\cal L}_{J=0} 
& = & 
\sum_{\rm{scalars}} \quad
\left(D_\mu\Phi\right)^{\dag} \left(D^\mu\Phi\right)
-
m^2 \Phi^{\dag}\Phi
\\
{\cal L}_{J=1} 
& = & 
\sum_{\rm{vectors}} \quad
-{1\over2} \left(D_\mu\Phi^\nu-D_\nu\Phi^\mu\right)^{\dag} 
           \left(D^\mu\Phi_\nu-D^\nu\Phi_\mu\right)
+
m^2 \Phi_\mu^{\dag}\Phi^\nu
\label{comp}
\end{eqnarray}
where $\Phi$ and $A$ are the \lk\ and photon fields,
$D_\mu = \partial_\mu - ieQA_\mu$
is the covariant derivative,
$e$ is the \EM\ coupling constant and
$m$ and $Q$ are the \lk\ mass and \EM\ charge.
This lagrangian describes the minimal vector boson coupling,
typical of a composite \lk.
If, however,
the vector \lk s are gauge bosons,
an extra Yang-Mills piece has to be added
in order to maintain gauge invariance:
\beq
{\cal L}_{G} =  
\sum_{\rm{vectors}} \quad
-ie ~\Phi_\mu^{\dag}\Phi_\nu \left(\partial^\mu A^\nu - \partial^\nu A^\mu\right)
\ .\label{ym}
\eeq
If this piece is not included,
tree-level unitarity is bound to be lost.
Indeed,
the effective lagrangian (\ref{comp})
is no longer valid at energies 
of the order of the scale 
where 
the full (gauge) theory from which it was derived 
has to be considered.

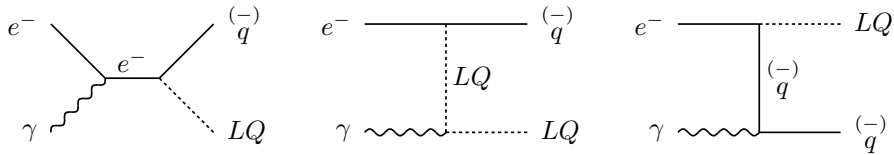
\begin{figure}[htb]
\hskip20mm\input{feyn.tex}
\bigskip
\caption{
  Lowest order Feynman diagrams for the production of \lk s.
}
\label{feyn}
\end{figure}

\begin{figure}[htb]
{\normalsize$\sigma$ [pb]}
\vskip26mm\hskip13mm
\epsfig{file=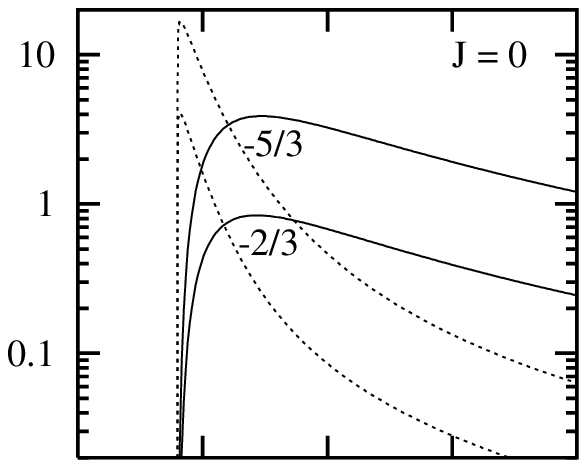,bbllx=4cm,bblly=4.2cm}
  \hskip-26mm
\epsfig{file=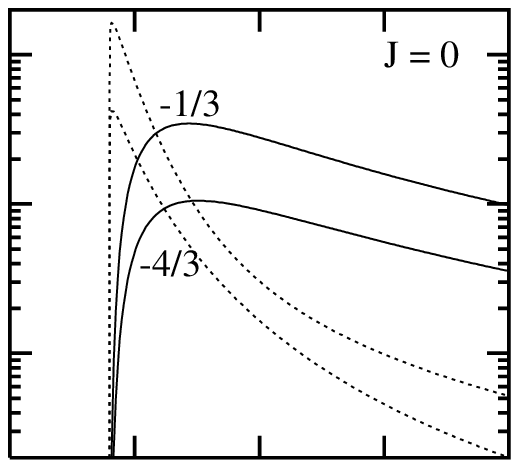,bbllx=-7cm,bblly=4.2cm}
\vskip42mm\hskip13mm
\epsfig{file=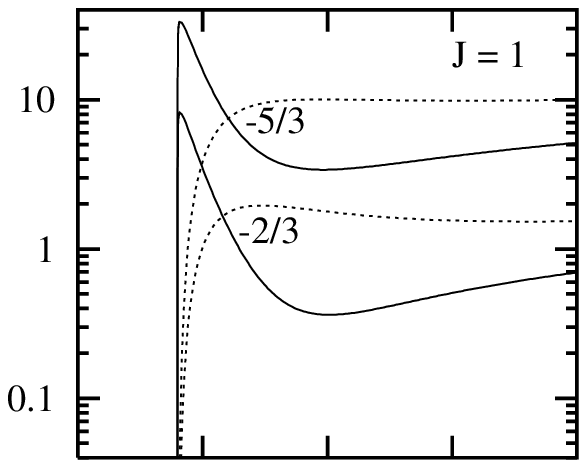,bbllx=4cm,bblly=4.2cm}
  \hskip-26mm
\epsfig{file=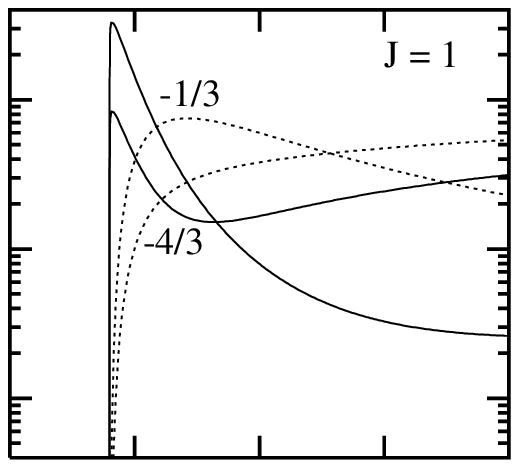,bbllx=-7cm,bblly=4.2cm}
\vskip42mm\hskip13mm
\epsfig{file=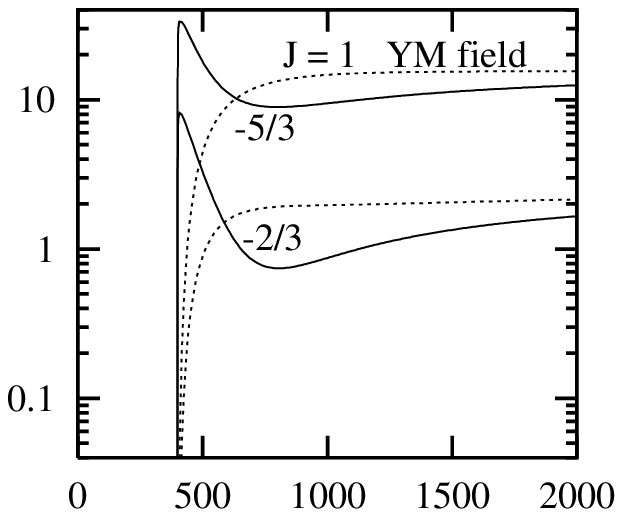,bbllx=4cm,bblly=4.2cm}
  \hskip-26mm
\epsfig{file=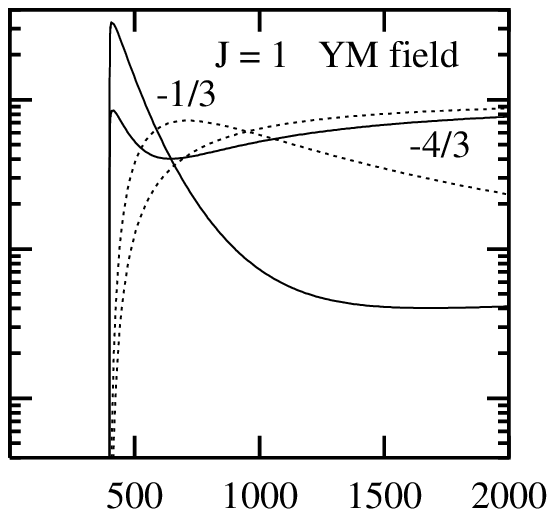,bbllx=-7cm,bblly=4.2cm}
\vskip24mm
\hskip30mm{\normalsize$\sqrt{s}$ [GeV]}\hskip35mm{\normalsize$\sqrt{s}$ [GeV]}
\vskip-1mm
\caption{
  Energy dependence of 400 GeV \lk\ production \xs s.
  The solid curves correspond to the choice 
  of respective electron and photon polarizations yielding $P_eP_\gamma=1$,
  whereas the dotted curves are for $P_eP_\gamma=-1$.
  The charges of the \lk s are indicated next to the corresponding curves.
}
\label{eny}
\end{figure}

Ignoring the resolved photon contributions
the typical $s$-, $t$- and $u$-channel Feynman diagrams
for the 
$e^-\gamma \to q LQ$ reaction
are shown in Fig.~\ref{feyn}.
There are 24 different types of processes,
depending on 
whether the produced \lk\ 
is a scalar, vector or gauge boson,
whether it couples to right- or left-handed leptons
and what is its \EM\ charge ($Q=-1/3,-2/3,-4/3,-5/3$). 
We display the energy behaviour
of the production \xs s
in Fig.~\ref{eny}.
For the purpose of these plots,
we have set the generic \lk-electron-quark coupling $\lambda$
equal to the \EM\ coupling constant
$\lambda=e$.
As in the rest of this report,
we also assume the electron and photon beams to be both
100 \%\ polarized and monochromatic.
A more detailed analysis,
with realistic polarizations and energy spectra \cite{ginzburg},
can be found in the last of Refs~\cite{eg}.

\clearpage
\thispagestyle{empty}
\clearpage

The threshold behaviour of the \xs s
mainly depends on the $u$-channel singularity
of the third graph in Fig.~\ref{feyn}
and on the electron and photon relative polarizations:
\beq
\left\{ \quad
\begin{array}{l}
  \displaystyle{\sigma(J=0) \quad \propto \quad ~(1+Q)^2 
    \quad (1 \mp P_\gamma) (1 \pm P_e) }
  \\\\
  \displaystyle{\sigma(J=1) \quad \propto \quad 2(1+Q)^2 
    \quad (1 \pm P_\gamma) (1 \pm P_e) }
\end{array}
\right.
\label{thresh}
\eeq
where $P_e$ and $P_\gamma$ 
are the polarizations of the electron and
photon beams.
The scalar \xs s quickly drop to zero 
for initial state electrons and photons with the same helicity,
whereas the vector \xs s are suppressed
when they have opposite helicities.
For equal couplings,
the vector threshold \xs s are twice as large as the scalar ones.

In the asymptotic region
the $J=0$ and $J=1$ \lk s also display very different behaviours.
Whereas the scalar \xs s decrease like 
$1/s$,
the vector \xs s eventually increase like 
$\ln s$.
If the vectors are gauge fields, though,
their \xs s saturate for large values of $s$. 

\begin{figure}[htb]
\centerline{\input{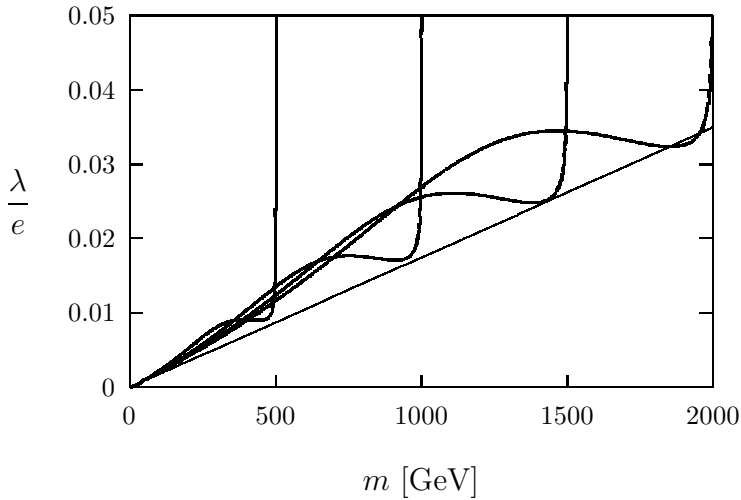}}
\bigskip
\caption{
  Loci of $\sigma(J=1,Q=-5/3)=1$ fb
  as a function of the \lk\ mass and coupling to fermions.
  The collider energies are from left to right .5, 1, 1.5 and 2 TeV
  and $P_eP_\gamma=1$.
  The thinner osculating line is given by Eq.~(\protect\ref{osc}).
}
\label{lim}
\end{figure}

In general,
the \lk s decay into a charged lepton and a jet
with a substantial branching ratio.
Even if the lepton is an electron,
the \sm\ background can easily be rendered harmless
by simple invariant mass cuts.
To estimate the \lk\ discovery potential
of \ep\ collisions,
we have plotted in Fig.~\ref{lim}
the boundary in the $(m_{LQ},\lambda/e)$ plane, 
below which the \xs\
of a $Q=-5/3$ Yang-Mills \lk\
yields less than 10 events.
For this we consider four different collider energies
and assume 10 fb$^{-1}$ of accumulated luminosity.
In general,
these curves are approximately osculated by the line
\beq
{\lambda \over e} \quad = \quad 0.02 
\quad {m/\mbox{TeV} \over \sqrt{{\cal L}/{\rm fb}^{-1}}}
\quad \sqrt{{n \over (J+1) (1+Q)^2 }}
\qquad
\left(
  m \le .9\sqrt{s}
\right)
\ ,\label{osc}
\eeq
where 
$\lambda$ is the \lk's coupling to leptons and quarks,
$m$ its mass,
$Q$ its charge,
$J$ its spin,
$n$ is the required number of events and
$\cal L$ is the available luminosity.
This scaling relation
provides a convenient means to gauge
the \lk\ discovery potential of \ep\ scattering.
It is valid for \lk\ masses short off 90\%\ of the collider energy
and assumes 
the electron and photon beams to be 100\%\ polarized
while their relative helicities are chosen 
such as to enhance the threshold \xs\ 
({\em cf.} Eqs~(\ref{thresh})).

In comparison,
the best bounds on the \lk\ couplings,
which have been derived indirectly from low energy data \cite{sacha},
are no better than
\beq
{\lambda \over e} \quad \ge \quad m/\mbox{TeV}
\ ,
\eeq
for \lk\ interactions involving only the first generation.
Similar bounds on couplings involving higher generations
are even poorer.

If a \lk\ is discovered someday,
it becomes interesting to determine its nature.
In principle,
\ep\ scattering may discriminate between the 24
combinations of the quantum numbers
$J$, $Q$ and $P$,
where the latter is the chirality of the electron to which the \lk\ couples.
For the case $J=1$ we make the distinction between 
gauge and non-gauge \lk s.

It is of course trivial to determine $P$
by switching the polarization of the electron beam.
Similarly,
it is almost as easy to distinguish scalars from vectors.
Indeed,
as can be inferred from Eqs~(\ref{thresh})
and from Fig.~\ref{eny},
all threshold \xs s are very sensitive 
to the relative electron and photon polarizations.
Since this effect works in opposite directions
for $J=0$ and $J=1$,
a simple photon polarization flip upon discovery
should suffice to determine
the spin of the discovered \lk s.

To determine the other quantum numbers,
we turn to the differential distributions.
As it turns out,
the interferences between the different channels
result into rather complex angular dependences of the \xs s.
There are even radiation zeros
for the reactions involving
scalar and Yang-Mills \lk s of charges -1/3 and -2/3.

\clearpage
\begin{figure}[htb]
\hskip3mm
\raisebox{2mm}{\normalsize$\displaystyle{1\over\sigma}{\d\sigma \over \d\theta}$}
\hskip20mm
\raisebox{-0mm}{\normalsize$\displaystyle P_eP_\gamma=1$}
\hskip35mm
\raisebox{-0mm}{\normalsize$\displaystyle P_eP_\gamma=-1$}
\vskip-8mm
\hskip3mm
\renewcommand{\arraystretch}{2.7}{\normalsize$\begin{array}[b]{@{}r@{}}\\10^{-2}\\10^{-3}\\10^{-4}\\10^{-5}\end{array}$}\renewcommand{\arraystretch}{1}\hskip5mm
\raisebox{18mm}{
\epsfig{file=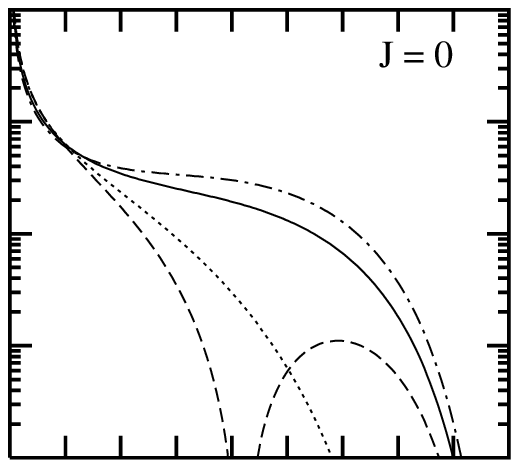,bbllx=4cm,bblly=4.2cm}
  \hskip-26mm
\epsfig{file=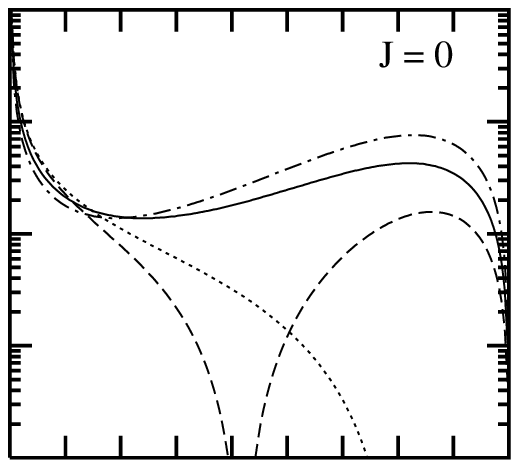,bbllx=-7cm,bblly=4.2cm}
}
\vskip-11mm
\hskip3mm
\renewcommand{\arraystretch}{2.7}{\normalsize$\begin{array}[b]{@{}r@{}}\\10^{-2}\\10^{-3}\\10^{-4}\\10^{-5}\end{array}$}\renewcommand{\arraystretch}{1}\hskip5mm
\raisebox{18mm}{
\epsfig{file=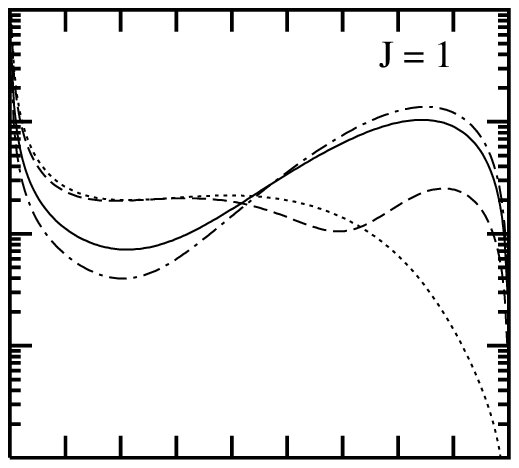,bbllx=4cm,bblly=4.2cm}
  \hskip-26mm
\epsfig{file=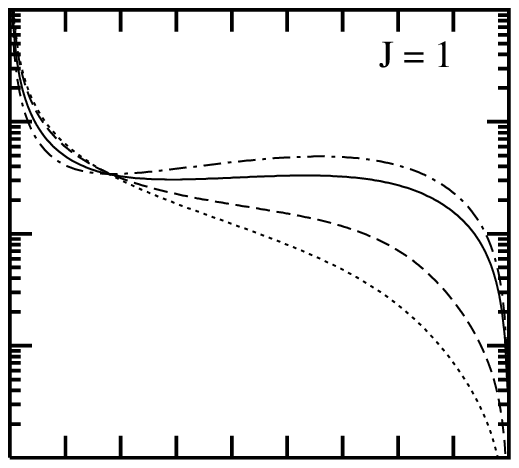,bbllx=-7cm,bblly=4.2cm}
}
\vskip-11mm
\hskip3mm
\renewcommand{\arraystretch}{2.7}{\normalsize$\begin{array}[b]{@{}r@{}}\\10^{-2}\\10^{-3}\\10^{-4}\\10^{-5}\end{array}$}\renewcommand{\arraystretch}{1}\hskip5mm
\raisebox{18mm}{
\epsfig{file=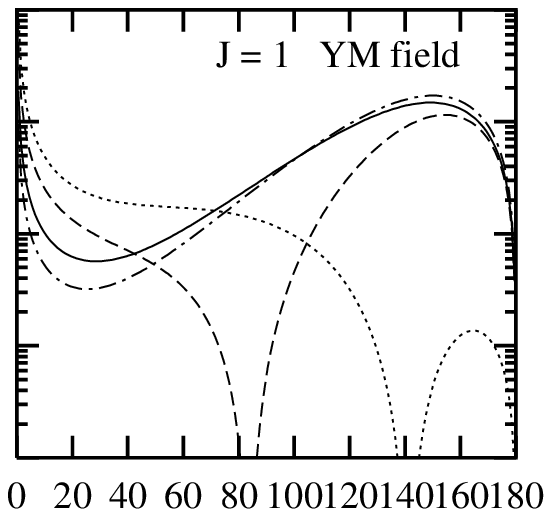,bbllx=4cm,bblly=4.2cm}
  \hskip-26mm
\epsfig{file=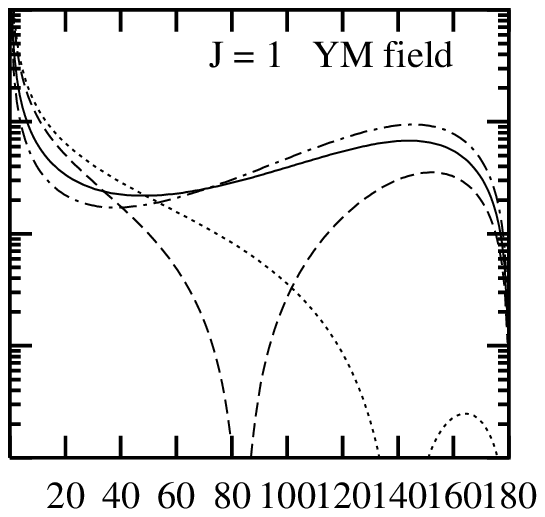,bbllx=-7cm,bblly=4.2cm}
}
\vskip3mm
\hskip3mm\hskip35mm{\normalsize$\theta$}\hskip50mm{\normalsize$\theta$}
\vskip-2mm
\caption{
  Angular distributions of 500 GeV \lk s produced in 1 TeV collisions.
  The coding of the curves is\quad
  solid: $Q=-5/3$;
  dot-dashed: $Q=-4/3$;
  dashed: $Q=-2/3$;
  dotted: $Q=-1/3$.
}
\label{ang}
\end{figure}
\clearpage

These very salient features
are displayed in Fig.~\ref{ang}
and could well be observed
away from threshold,
where the $u$-channel pole is no longer so dominant.
In this figure,
we have considered 500 GeV \lk s produced at a 1 TeV collider.

\newcommand{\rb}[1]{\raisebox{2ex}[-2ex]{$#1$}}
\renewcommand{\arraystretch}{1.5}
\begin{table}[htb]
$$
\begin{array}{|c|c||c|c|c|c|}
\hline
\multicolumn{2}{|c||}{}
& \multicolumn{4}{|c|}{J=0}
\\\cline{3-6}
\multicolumn{2}{|c||}
{\rb{\mbox{\boldmath$\displaystyle\Large {\lambda \over e}\sqrt{{\cal L}/{\rm fb}^{-1}} $}}}
& {\tiny -{5\over3}} & {\tiny -{2\over3}} & {\tiny -{4\over3}} & {\tiny -{1\over3}} 
\\\hline\hline
& {\tiny -{5\over3}} & \times & .2 & .6 & .2
\\\cline{2-6}
& {\tiny -{2\over3}} & .6 & \times & .5 & 2.1
\\\cline{2-6}
\rb{J=0} & {\tiny -{4\over3}} & .9 & .2 & \times & .2
\\\cline{2-6}
& {\tiny -{1\over3}} & .3 & 1.0 & .2 & \times
\\\hline
\end{array}
$$
\vskip-2mm
$$
\begin{array}{|c|c||c|c|c|c|c|c|c|c|}
\hline
\multicolumn{2}{|c||}{} & \multicolumn{4}{|c|}{J=1} & \multicolumn{4}{|c|}{J=1\quad\mbox{YM field}}
\\\cline{3-10}
\multicolumn{2}{|c||}
{\rb{\mbox{\boldmath$\displaystyle\Large {\lambda \over e}\sqrt{{\cal L}/{\rm fb}^{-1}} $}}}
 & {\tiny -{5\over3}} & {\tiny -{2\over3}} & {\tiny -{4\over3}} & {\tiny -{1\over3}} & {\tiny -{5\over3}} & {\tiny -{2\over3}} & {\tiny -{4\over3}} & {\tiny -{1\over3}} 
\\\hline\hline
& {\tiny -{5\over3}} & \times & .2 & .7 & .2 & .3 & .2 & .2 & .1
\\\cline{2-10}
& {\tiny -{2\over3}} & .6 & \times & .6 & 1.0 & .4 & .4 & .3 & .6
\\\cline{2-10}
\rb{J=1} & {\tiny -{4\over3}} & 1.0 & .3 & \times & .2 & .5 & .3 & .4 & .2
\\\cline{2-10}
& {\tiny -{1\over3}} & .3 & .6 & .2 & \times & .2 & .2 & .2 & .8
\\\hline
& {\tiny -{5\over3}} & .2 & .07 & .2 & .07 & \times & .2 & .7 & .06
\\\cline{2-10}
\rb{J=1} & {\tiny -{2\over3}} & .6 & .4 & .5 & .3 & .7 & \times & .6 & .3
\\\cline{2-10}
\rb{\mbox{YM}} & {\tiny -{4\over3}} & .2 & .09 & .2 & .09 & 1.0 & .2 & \times & .08
\\\cline{2-10}
\rb{\mbox{field}} & {\tiny -{1\over3}} & .2 & .4 & .2 & .9 & .2 & .2 & .2 & \times
\\\hline
\end{array}
$$
\caption{
Smallest values of
$\lambda/e\protect\sqrt{{\cal L}/{\rm fb}^{-1}}$
which allow discriminating 
with 99.9\%\ confidence
two different 500 GeV scalar or vector \lk s
at a 1 TeV collider.
Scalar and vector \lk s
can easily be told apart
by flipping the laser polarization at threshold.
The charge of the \lk s is listed in the second row and column.
The \lk s listed in the rows 
are assumed to produce the data,
whereas those listed in the columns 
are tested against this data.
}
\label{ts}
\end{table}

To roughly estimate the \ep\ potential
for discriminating the different types of \lk s,
we compare these differential distributions
with a Kolmogorov-Smirnov test.
Focusing on the case of Fig.~\ref{ang},
where a 500 GeV \lk\ is studied at a 1 TeV collider,
Tables~\ref{ts}
summarize the minimal values of
$$
{\lambda \over e} \quad \sqrt{{\cal L}/{\rm fb}^{-1}}
$$
needed to tell apart 
two different types of \lk s
with 99.9\%\ confidence.
Assuming each combination of electron and photon polarizations
has accumulated 50 fb$^{-1}$ of data,
some \lk\ types have so different angular distributions
that a coupling as small as $\lambda=.0085e$ 
is quite sufficient to tell them apart.
Others need as much as $\lambda=.3e$.
Obviously,
these numbers are only valid for this particular choice 
of \lk\ mass and collider energy.
They could be improved with a more sophisticated analysis
but should be indicative of what resolving power
an \ep\ scattering experiment can achieve.

To summarize,
the advantage of an \ep\ experiment over,
{\em e.g.},
\pe\ annihilation
is that the \lk s need not be produced in pairs,
hence higher masses can be explored.
Moreover,
the spin of the \lk s can easily be determined
at threshold
by inverting the polarization of the photon beam.
Finally,
the prospects for discriminating 
different \lk s of the same spin
can be greatly enhanced
by studying angular correlations
away from threshold.

\medskip
It is a pleasure to thank 
Sacha Davidson, 
Slava Ilyin,
David London and
H\'el\`ene Nadeau
for their invaluable inputs.

\end{document}

%% file: feyn.tex
{\unitlength.5mm\SetScale{1.4185}
\unitlength.45mm\SetScale{1.27665}

\begin{picture}(90,40)(-15,0)
\Text(-4,32)[rc]{$e^-$}
\Text(-4,00)[rc]{$\gamma$}
\Line(00,32)(16,16)
\Line(16,16)(32,16)
\Line(32,16)(48,32)
\DashLine(32,16)(48,00){1}
\Photon(00,00)(16,16){1}{4}
\Text(52,32)[lc]{$\stackrel{\scriptsize(-)}{q}$}
\Text(52,00)[lc]{$LQ$}
\Text(24,18)[cb]{$e^-$}
\end{picture}
\begin{picture}(90,40)(-15,0)
\Text(-4,32)[rc]{$e^-$}
\Text(-4,00)[rc]{$\gamma$}
\Line(00,32)(24,32)
\Line(24,32)(48,32)
\DashLine(24,32)(24,00){1}
\DashLine(24,00)(48,00){1}
\Photon(00,00)(24,00){1}{4}
\Text(52,32)[lc]{$\stackrel{\scriptsize(-)}{q}$}
\Text(52,00)[lc]{$LQ$}
\Text(26,16)[lc]{$LQ$}
\end{picture}
\begin{picture}(0,40)(-15,0)
\Text(-4,32)[rc]{$e^-$}
\Text(-4,00)[rc]{$\gamma$}
\Line(00,32)(24,32)
\Line(24,32)(24,00)
\Line(24,00)(48,00)
\DashLine(24,32)(48,32){1}
\Photon(00,00)(24,00){1}{4}
\Text(52,32)[lc]{$LQ$}
\Text(52,00)[lc]{$\stackrel{\scriptsize(-)}{q}$}
\Text(26,16)[lc]{$\stackrel{\scriptsize(-)}{q}$}
\end{picture}

}